# A silicon-based device for dynamic control of thermal emission

Sid Assawaworrarit,[1,*] Alex Song,[1] and Shanhui Fan[2]

[1]School of Electrical and Computer Engineering, Faculty of Engineering, The University of Sydney, NSW, Australia

[2]Department of Electrical Engineering, Ginzton Laboratory, Stanford University, Stanford, California, USA

*Correspondence: sid.assa@sydney.edu.au

Abstract:

Control of thermal emission is important in a number of applications from thermal energy harvesting and management and sensing of gas and chemical to thermal camouflage. Semiconductor-based devices can be engineered to enable electrical control of thermal emission, offering high modulation speed and ease of voltage control. Existing device designs for modulating thermal emission rely on semiconductors other than silicon, such as III-V and II-VI compounds, which are expensive. The silicon platform offers several advantages, including significantly lower cost, CMOS compatibility, and mature fabrication processes. However, a silicon-based design for modulating thermal emission remains absent. Here, we present an all-silicon device utilising electrical control over carrier dynamics to modulate a narrowband thermal emission in the mid-infrared region. We design a silicon device exhibiting voltage-controlled narrowband thermal emission at 10 μm and confirm its performance using electromagnetic calculations. This work paves the way for scalable, low-cost, and integrated thermal emission devices made possible by the silicon platform.

**Introduction**

Thermal radiation is a fundamental phenomenon, with objects at typical terrestrial temperatures emitting energy predominantly in the mid-infrared region. The ability to engineer and control this emission is critical for a wide range of applications[1–4], including energy conversion[5–7], thermal management[8–10], chemical and gas sensing[11,12], and thermal camouflage[13–19]. Recent advances in nanophotonic materials have enabled unprecedented control over thermal radiation properties and transformed the engineering of thermal emission control from conventional thermal emitters to systems capable of exhibiting coherent, narrowband, and directional emission[1–4,20–27]. Concurrently, there has been growing research interest in achieving dynamic control over thermal radiation[3,4,13,16,22,24,28–32], a capability that enables real-time modulation of thermal emission.

Two broad pathways exist to enable dynamic control of thermal radiation. The spectral radiance of the thermal radiation from an object can be expressed as $E(\lambda, \Omega)I_{bb}(\lambda, T)$, where $E(\lambda, \Omega)$ is the object's spectral emissivity and represents a measure of its ability to emit radiation at a specific wavelength, $\lambda$, and solid angle, $\Omega$, and $I_{bb}(\lambda, T)$ is the spectral radiance of a blackbody at the same wavelength and temperature, $T$, as described by Planck's law of blackbody radiation. A straightforward pathway to modulate thermal emission is to actively heat or cool the object, thereby changing its temperature. However, this method is limited in its performance due to the generally large thermal capacitance of most materials, which results in high energy consumption and slow response times. An alternative pathway is to modulate the object's emissivity, which can be achieved through various strategies. These include the use of phase-change materials, which alter their optical properties upon undergoing an insulator-metal phase transition[14,30]; micro electromechanical (MEM) structures[33,34]; and modulation of carrier concentration in graphene[16,24] and semiconductors[22,28,29,31], which can be achieved via electrical biasing. Among these, modulation of carrier concentration in semiconductors has shown great promise due to its fast response times, low energy consumption, and ability to deliver significant changes in emissivity.

Current demonstrations of dynamic thermal emission via carrier concentration modulation in semiconductors primarily rely on high-cost semiconductor platforms such as gallium arsenide (GaAs), as demonstrated by the Noda group[28,29] and Greffet group[22], and indium arsenide (InAs), as demonstrated by the Brongersma group[31]. However, silicon, which is abundant and significantly more cost-effective by a factor of over 100 compared to gallium arsenide[35], has been largely overlooked due to its inherently low emissivity in the mid-infrared wavelength range. Despite these challenges, engineering such a device on silicon offers advantages. Silicon's widespread availability and well-established and scalable fabrication processes make it ideal for large-scale integration and cost-effective manufacturing. Additionally, silicon-based platforms enable compatibility with existing semiconductor fabrication infrastructure.

In this paper, we demonstrate through device design calculations that dynamic control of thermal emissions can be achieved in an all-silicon platform. Our device operates by manipulating the carrier concentration within a silicon p-n junction, which is controlled using an applied bias voltage. Through detailed device simulations, we show that thermal emission can be dynamically modulated by applying voltage bias. When combined with a tailored photonic structure, this approach enables narrowband thermal emission characteristics. By demonstrating the feasibility of silicon-based systems for dynamic thermal control, this work paves the way for scalable and cost-effective technology in mid-infrared dynamic thermal emitter design.

# Results

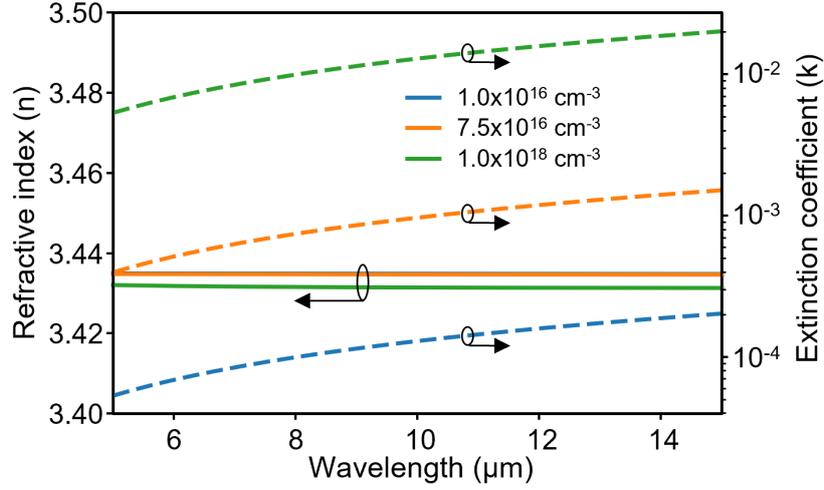

Figure 1| Complex refractive indices ($n + ik$) of n-doped silicon in the mid-infrared range, calculated using the Drude model for the indicated free carrier concentrations.

**Modelling carrier-induced thermal emissivity in silicon**. For an object obeying Lorentz reciprocity, the emissivity and absorptivity are equal as dictated by the Kirchhoff's law [36]. The Kichhoff's law enables us to obtain a device's emissivity by calculating its absorptivity. In the mid-infrared region, silicon is relatively transparent with free carriers dominating the optical absorption process[37]. The optical response arising from free carrier dynamics can be modelled using the Drude model, which gives a frequency-dependent dielectric function[38],

$$\epsilon(\omega) = \epsilon_\infty - \frac{\omega_p^2}{\omega^2 + i\omega\gamma},$$

where $\epsilon_\infty$ is the high-frequency dielectric constant, $\omega_p$ is the plasma frequency, and $\gamma$ is the damping rate. The plasma frequency is related to the carrier concentration, $N$, via $\omega_p^2 = \frac{Ne^2}{\epsilon_0 m^*}$ where $e$ is the elementary charge, $\epsilon_0$ is the vacuum permittivity, and $m^*$ is the effective mass of the charge carrier. The damping rate $\gamma$ is related to the carrier mobility, $\mu$, via $\gamma = \frac{e}{m^*\mu}$.

Experimental data show agreement between the Drude model and measured absorptivity values[37,39]. Figure 1 shows the complex refractive indices of doped silicon in the mid-infrared region for varying carrier concentrations, calculated using the Drude model with representative parameters for doped polysilicon from ref. [40]: $\epsilon_\infty = 11.8$, $m^* = 0.26 m_e$ for holes and $m^* = 0.36 m_e$ for electrons, where $m_e$ is the electron mass, and $\mu = 10$ cm$^2$/Vs. Under moderate doping levels, doped silicon behaves as a dielectric material, with absorption loss increasing with carrier concentration.

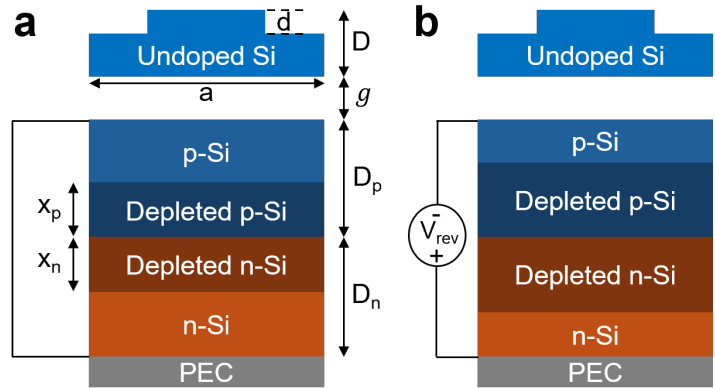

Figure 2| Device design and operation. (a) Schematic of the silicon p-n junction structure showing the initial carrier distribution. The p-n junction structure is separated from a photonic crystal slab by an air gap. The photonic crystal slab is made of undoped silicon and is not subject to electrical bias. (b) Under reverse bias, the carrier-depleted region expands, modulating the device's thermal emission. The dimensions used for device calculation in this paper are as follows: $a = 8.9$ μm, $D_p = D_n = 250$ nm, $D = 40$ nm, $g = 14$ nm, $d = 16$ nm.

**Device design and operation of a voltage-controlled thermal emitter in silicon.** To demonstrate that carrier-induced absorption and emission processes in silicon can enable a voltage-controlled thermal emitter, we propose a representative device architecture, design a specific example, and validate its performance through electromagnetic calculations.

Our device architecture, shown in Figs. 2a and b, employs a guided-mode resonance structure to realise voltage-controlled narrowband thermal emission from free carriers in silicon. A detailed analysis of guided-mode resonance and its application to narrowband thermal emission can be found in ref.[41]. The architecture comprises a top photonic crystal slab, separated by an air gap from a silicon p-n junction, with a perfect electric conductor (PEC) acting as a back reflector. This structure supports guided resonance modes that couple to free-space radiation from above the top surface of the photonic crystal. The operation of the device proceeds as follows: The carriers in p-n junction generates thermal emission, which can be controlled by the applied voltage bias. This emission couples into the guided resonance mode within the plane of the device. The air gap serves dual purposes: it provides electrical isolation between the p-n junction and the photonic crystal, and it enables tuning of the device's optical response. The top photonic crystal facilitates coupling between the guided modes and free-space radiation by periodically scattering the guided modes, which propagate along the slab plane, out to free space. This external coupling is most effective under a phase-matching condition, where the propagation constant of the guided mode matches with the periodicity of the photonic crystal, establishing a guided resonance[42]. The outcoupling of thermal emission via guided resonance modes results in narrowband thermal emission at a wavelength determined by the periodicity and geometry of the photonic crystal.

Achieving unity absorptivity and emissivity at the guided resonant wavelength also requires a balance between the rate at which the guided resonance modes are thermally generated and the external coupling rate at which they outcouple to free space. The internal generation rate of these modes depends on the carrier concentration within the p-n junction and its spatial distribution relative to the mode's electromagnetic field profile. Under this critical coupling condition, the device realises unity peak absorptivity and emissivity at the resonant wavelength, as shown in ref.[41].

Applying a negative voltage bias to the p-n junction depletes the carriers in the junction (Fig. 2b), thereby reducing the internal generation rate. This occurs because the depletion region expands, diminishing the number of free carriers available to interact with the electromagnetic field of the guided resonance mode. If the device is designed to achieve unity peak emissivity at zero bias, the reduction in internal generation under bias conditions lowers both absorptivity and emissivity. This applied voltage-induced mismatch between internal and external coupling rates enables voltage-controlled modulation of thermal emission.

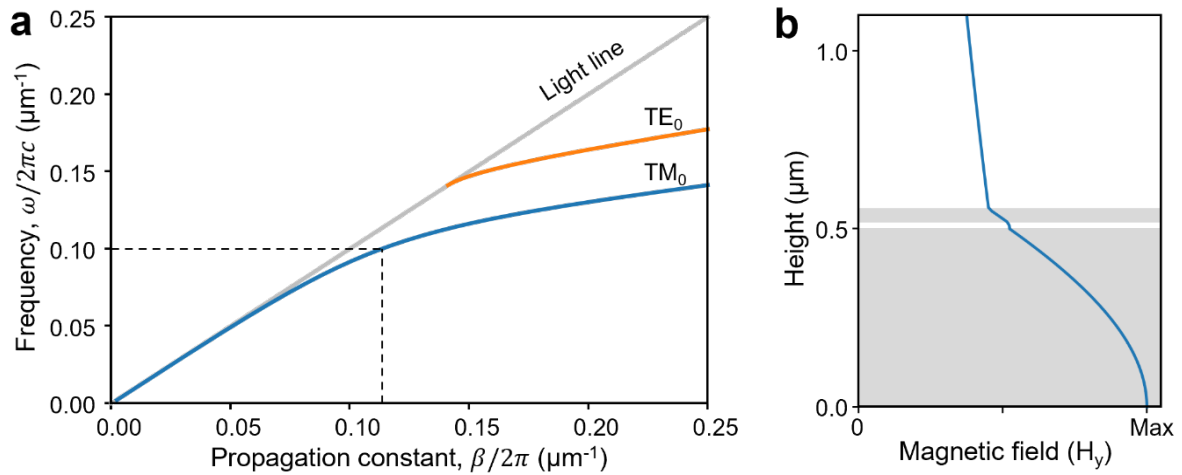

Figure 3| Guided mode resonance analysis. (a) Dispersion relation of guided modes in a uniform unpatterned device structure. Dashed lines indicate the target mode at 10 μm. (b) Fundamental transverse magnetic ($TM_0$) mode with out-of-plane magnetic field ($H_y$) profile at 10 μm. Shaded regions denote silicon.

To demonstrate a concrete device design and its performance characteristics, we present a specific silicon device implementation that achieves narrowband, polarisation-selective, and voltage-controllable emissivity at a wavelength of 10 μm. The design uses a p-n junction with equal doping concentrations on both sides of the junction: $N_A = N_D = 7.5 \times 10^{16}$ cm$^{-3}$, for the p- and n-side respectively, and equal thicknesses of $D_p = D_n = 250$ nm for the p- and n-type layers. The silicon photonic crystal slab has a thickness of $D = 40$ nm. These dimensions are chosen to be sufficiently thin, with $D_p + D_n + D < \lambda/4n$ where $\lambda = 10$ μm is the target wavelength and $n$ is silicon's refractive index, so that the structure supports a single guided mode at 10 μm to ensure a well-separated narrowband emissivity response. Additionally, a substantial portion of the structure is dedicated to the p–n junction to maximise the influence of carrier modulation on the device's emissivity response. The dispersion relation, calculated using guided-mode analysis[43] and shown in Fig. 3a, confirms this single-mode behaviour for the $TM_0$ (magnetic field out of plane) mode at 10 μm, while the $TE_0$ (electric field out of plane) mode has its cutoff frequency at a shorter wavelength (higher frequency). The field profile of the $TM_0$ mode at 10 μm, shown in Fig. 3b, shows a substantial mode overlap with the p–n junction region, ensuring that carrier modulation in the junction results in emissivity changes. Next, a one-dimensional grating with a period of 8.9 μm with a 50% duty cycle is etched into the top surface of the floating slab. The one-dimensional grating structure is chosen for its simplicity, with its period chosen to satisfy the phase-matching condition for the $TM_0$ mode at 10 μm, thereby enabling narrowband emissivity at that wavelength.

**Device emissivity calculation.** To compute the emissivity of the device, we employ the rigorous coupled-wave analysis (RCWA) method, a numerical technique for solving Maxwell's equations in periodic structures[43,44]. This method enables efficient computation of optical scattering parameters by considering the field in a single unit cell of the periodic structure and solving the problem in the frequency domain. RCWA is well-suited for our device configuration (Figs. 2a and 2b), which comprises a periodic multilayer structure featuring a photonic crystal silicon slab and a graded dielectric profile resulting from varying carrier concentrations in the p-n junction region.

To perform RCWA calculations for our device design, we divide the device into distinct layers, each characterised by a unique dielectric function based on material composition, carrier type, and carrier concentration. For the undoped silicon photonic crystal slab, a constant dielectric function $\epsilon = 11.8$ is assigned. For the p-n junction region, we further subdivide the structure into four layers, which are, from top to bottom, as follows: (i) p-doped silicon with a high hole concentration of $N_A$, (ii) a depletion layer of p-doped silicon with negligible carrier concentration, (iii) a depletion layer of n-doped silicon with negligible carrier concentration, and (iv) n-doped silicon with a high electron concentration of $N_D$. Each of these regions is assigned a dielectric function using the Drude model (Eq. 1), which captures the frequency-dependent optical response of free carriers in doped semiconductor materials.

The thicknesses of the p- and n-side depletion regions are determined by the applied voltage bias, and can be calculated using the abrupt junction model[45] as $x_p = \sqrt{\frac{2\epsilon_{Si}(V_{bi}+V_{rev})}{N_A\left(1+\frac{N_A}{N_D}\right)}}$ and $x_n = \sqrt{\frac{2\epsilon_{Si}(V_{bi}+V_{rev})}{N_D\left(1+\frac{N_D}{N_A}\right)}}$, respectively, where $\epsilon_{Si}$ is the permittivity of silicon and $V_{bi}$ is the built-in voltage for the junction.

We use an open-source RCWA package[44] to compute the device's reflectivity. The device's emissivity is then derived from the difference between the reflectivity and unity.

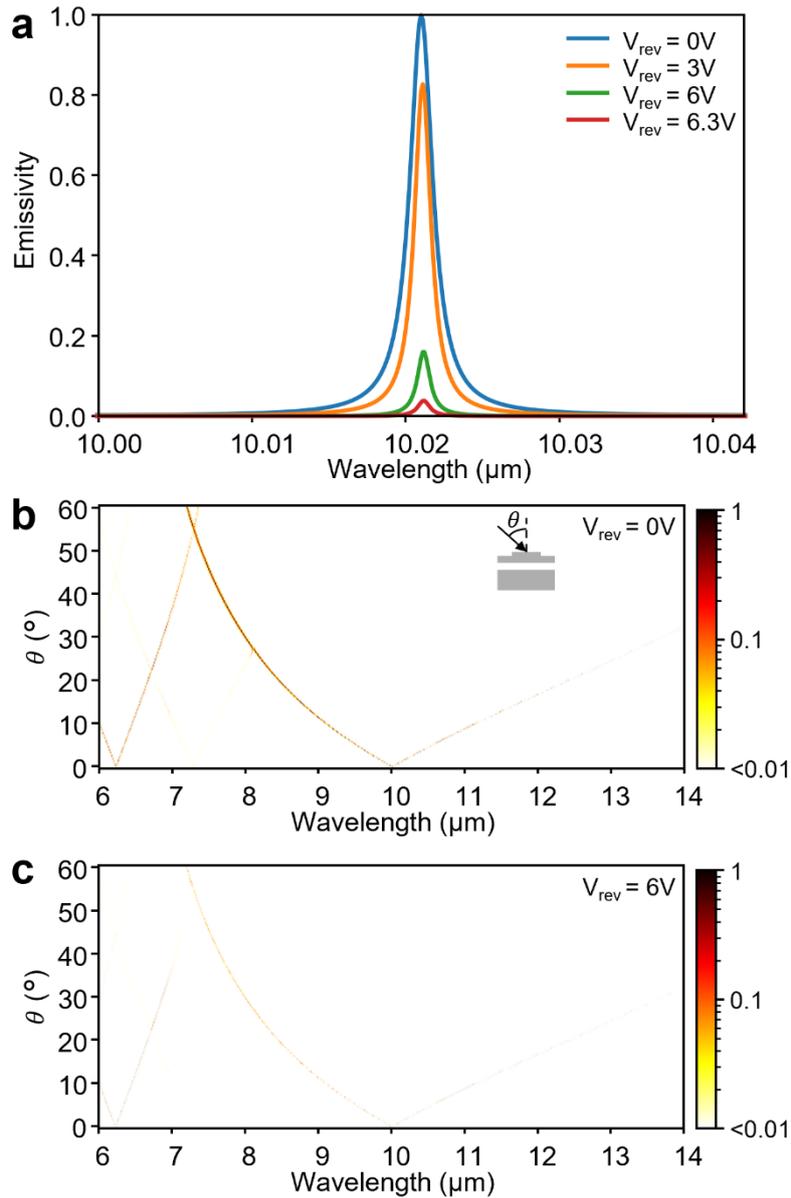

Figure 4| Calculated device mid-infrared emissivity. (a) Calculated emissivity spectra under normal incidence for TM polarisation at various applied voltages. (b), (c) Calculated emissivity maps at oblique incidence angles for 0 V (b) and -6 V (c) applied voltages.

**Voltage control of thermal emission**. The calculated emissivity spectra of the device under various applied voltages in the normal incidence direction are shown in Fig. 4a for TM polarisation. (The emissivity for TE polarisation is minimal (<0.1%) for all applied voltages.) The device exhibits voltage-tuneable, narrowband emissivity centred at 10μm for the TM polarisation. To achieve unity peak emissivity at the design wavelength for TM polarisation under zero applied voltage, the air gap ($g$) and grating depth ($d$) were carefully optimised. The full-width-at-half-maximum (FWHM) linewidth of the emissivity peak is 2 nm.

When a reverse bias voltage is applied, the device emissivity is suppressed due to the depletion of charge carriers in the p- and n-regions of the p-n junction. At a reverse bias of 6.3 V, most carriers are removed from the junction, leading to a reduction in the emissivity peak to ~0.04.

In contrast, the TE polarisation exhibits virtually zero emissivity across all applied voltages. This is due to the absence of guided resonance for the TE polarisation at 10 μm, as this wavelength lies below the cutoff frequency for the TE guided mode in the structure (see Fig. 3b).

For off-normal incidence, the emissivity spectra for the zero-bias voltage, shown in Fig. 4b, exhibit wavelength-angle-dependent behaviour, with the peak absorptivity shifting as the incident angle varies. The resonance at 10 μm splits into two distinct modes as the incident angle increases from normal. This angular dependence of the thermal emission wavelength results in the emission spectrum shifts across the mid-infrared range depending on the output angle. This effect can be understood from the dispersion relation of the guided mode near the grating wavevector. As the incident angle deviates from normal, the phase-matching condition for guided resonance shifts, causing the resonance to occur at different wavelengths. The application of reverse bias voltage suppresses emissivity across all incident angles, mirroring the behaviour observed under normal incidence as shown in Fig. 4c.

**Discussion**

We have designed a silicon-based device that uses the electrical control of charge carriers in a p-n junction to demonstrate electrical modulation of thermal emission. This work highlights silicon as a promising platform for modulating mid-infrared thermal radiation through carrier dynamics. Here, we discuss potential improvements and design considerations to enhance the device's performance.

Our device exhibits a narrowband thermal emission at a centre wavelength of 10 μm. This centre wavelength is a consequence of the periodicity of the photonic crystal structure, which can be engineered to achieve the desired emission wavelength. Unlike designs that rely on material-specific resonance, this approach allows for far greater flexibility in tuning the thermal wavelength across a wide range. The tunability of this design is particularly advantageous for applications requiring spectral control, such as spectroscopy and sensing.

The narrowband thermal emission of 2 nm can be advantageous for applications requiring high spectral selectivity, such as spectroscopy. The device design inherently allows tuning of the linewidth to meet application-specific requirements. For instance, we can broaden the linewidth of our device by increasing the doping concentration in both the p- and n-regions of the p-n junction, thereby increasing thermal emission from carriers, together with a corresponding increase in the grating depth to maintain critical coupling condition. Conversely, reducing doping with the corresponding reduction in grating depth narrows the linewidth.

The polarisation selectivity of our device, which emits narrowband thermal radiation only in the TM polarisation, is a consequence of the design of the top one-dimensional grating. This grating supports only TM-polarised guided resonance at 10 μm. This polarisation selectivity could be beneficial in polarisation-sensitive sensing applications. For polarisation-insensitive design, one can design a two-dimensional (2D) photonic crystal slab that, by symmetry, support both TM and TE polarisations at the same wavelength, enabling thermal emission in both polarisations.

The angular dependence of our device's thermal emission is also consequence of the photonic crystal design, specifically the dispersion of the guided mode near the grating wavevector. This angular could be useful for applications requiring directional or steerable thermal emission, such as directional sources with angularly modulated output. When angular independence is desired, the dispersion of the guided mode can be made flat by designing a 2D photonic crystal with a flat band structure, as demonstrated in ref.[46]. This eliminates the angular dependence of the guided resonance frequency, allowing the device to emit uniformly in all directions.

Our design relies on a floating photonic crystal slab above a p-n junction to provide electrical isolation as well as for tuning the emissivity response. Such a structure can be achieved by a MEM system[47], where the slab is supported by flexible anchors to enable precise control over the slab's position, or by putting support elsewhere[48].

Finally, silicon device designs that enable electrical control of thermal emission in silicon can be extended beyond the specific p-n junction realisation demonstrated here. Other silicon-based structures could offer advantages. For example, a design based on a metal-oxide-semiconductor (MOS) capacitor, which provides similar voltage control over carrier concentration, can be engineered to achieve voltage-controlled dynamic thermal emission.

**Conclusion**

We have presented an all-silicon device design that enables voltage control of thermal emission. The design integrates the thermal emission from carriers in a silicon p–n junction, whose quantity can be controlled by an applied voltage, with a photonic crystal to generate polarised, narrowband thermal emission in the mid-infrared region. The use of silicon offers great benefits in terms of cost-effectiveness, scalability, matured and established fabrication processes, and seamless integration with CMOS technology. This work opens new avenues for low-cost thermal emission management technologies with broad implications for mid-infrared sensing, thermal energy harvesting and management, and mid-infrared optoelectronics.

**Acknowledgements**

S.A. acknowledges the Horizon Fellowship from the University of Sydney. S. F. acknowledges the support of a MURI project from the U. S. Air Force Office of Scientific Research (Grant No. FA9550-21-1-0312)

References:
(1) Li, W.; Fan, S. Nanophotonic Control of Thermal Radiation for Energy Applications [Invited]. *Opt. Express* **2018**, *26* (12), 15995–16021. https://doi.org/10.1364/OE.26.015995.
(2) Baranov, D. G.; Xiao, Y.; Nechepurenko, I. A.; Krasnok, A.; Alù, A.; Kats, M. A. Nanophotonic Engineering of Far-Field Thermal Emitters. *Nat. Mater.* **2019**, *18* (9), 920–930. https://doi.org/10.1038/s41563-019-0363-y.
(3) Picardi, M. F.; Nimje, K. N.; Papadakis, G. T. Dynamic Modulation of Thermal Emission—A Tutorial. *J. Appl. Phys.* **2023**, *133* (11), 111101. https://doi.org/10.1063/5.0134951.
(4) Chu, Q.; Zhong, F.; Shang, X.; Zhang, Y.; Zhu, S.; Liu, H. Controlling Thermal Emission with Metasurfaces and Its Applications. *Nanophotonics* **2024**, *13* (8), 1279–1301. https://doi.org/10.1515/nanoph-2023-0754.
(5) De Zoysa, M.; Asano, T.; Mochizuki, K.; Oskooi, A.; Inoue, T.; Noda, S. Conversion of Broadband to Narrowband Thermal Emission through Energy Recycling. *Nat. Photonics* **2012**, *6* (8), 535–539. https://doi.org/10.1038/nphoton.2012.146.
(6) Assawaworrarit, S.; Zhou, M.; Fan, L.; Fan, S. Nighttime Electric Power Generation at a Density of 350 mW/M2 via Radiative Cooling. *Cell Rep. Phys. Sci.* **2025**, *6* (1). https://doi.org/10.1016/j.xcrp.2024.102362.
(7) Hassan, S.; Doiron, C. F.; Naik, G. V. Optimum Selective Emitters for Efficient Thermophotovoltaic Conversion. *Appl. Phys. Lett.* **2020**, *116* (2), 023903. https://doi.org/10.1063/1.5131367.
(8) Zhou, J.; Chen, T. G.; Tsurimaki, Y.; Hajj-Ahmad, A.; Fan, L.; Peng, Y.; Xu, R.; Wu, Y.; Assawaworrarit, S.; Fan, S.; Cutkosky, M. R.; Cui, Y. Angle-Selective Thermal Emitter for Directional Radiative Cooling and Heating. *Joule* **2023**, *7* (12), 2830–2844. https://doi.org/10.1016/j.joule.2023.10.013.
(9) Zhao, B.; Hu, M.; Ao, X.; Chen, N.; Pei, G. Radiative Cooling: A Review of Fundamentals, Materials, Applications, and Prospects. *Appl. Energy* **2019**, *236*, 489–513. https://doi.org/10.1016/j.apenergy.2018.12.018.
(10) Zhao, D.; Aili, A.; Zhai, Y.; Xu, S.; Tan, G.; Yin, X.; Yang, R. Radiative Sky Cooling: Fundamental Principles, Materials, and Applications. *Appl. Phys. Rev.* **2019**, *6* (2), 021306. https://doi.org/10.1063/1.5087281.
(11) Lochbaum, A.; Fedoryshyn, Y.; Dorodnyy, A.; Koch, U.; Hafner, C.; Leuthold, J. On-Chip Narrowband Thermal Emitter for Mid-IR Optical Gas Sensing. *ACS Photonics* **2017**, *4* (6), 1371–1380. https://doi.org/10.1021/acsphotonics.6b01025.
(12) Lochbaum, A.; Dorodnyy, A.; Koch, U.; Koepfli, S. M.; Volk, S.; Fedoryshyn, Y.; Wood, V.; Leuthold, J. Compact Mid-Infrared Gas Sensing Enabled by an All-Metamaterial Design. *Nano Lett.* **2020**, *20* (6), 4169–4176. https://doi.org/10.1021/acs.nanolett.0c00483.
(13) Xiao, L.; Ma, H.; Liu, J.; Zhao, W.; Jia, Y.; Zhao, Q.; Liu, K.; Wu, Y.; Wei, Y.; Fan, S.; Jiang, K. Fast Adaptive Thermal Camouflage Based on Flexible VO2/Graphene/CNT Thin Films. *Nano Lett.* **2015**, *15* (12), 8365–8370. https://doi.org/10.1021/acs.nanolett.5b04090.
(14) Qu, Y.; Li, Q.; Du, K.; Cai, L.; Lu, J.; Qiu, M. Dynamic Thermal Emission Control Based on Ultrathin Plasmonic Metamaterials Including Phase-Changing Material GST. *Laser Photonics Rev.* **2017**, *11* (5), 1700091. https://doi.org/10.1002/lpor.201700091.
(15) Li, Y.; Bai, X.; Yang, T.; Luo, H.; Qiu, C.-W. Structured Thermal Surface for Radiative Camouflage. *Nat. Commun.* **2018**, *9* (1), 273. https://doi.org/10.1038/s41467-017-02678-8.

(16) Salihoglu, O.; Uzlu, H. B.; Yakar, O.; Aas, S.; Balci, O.; Kakenov, N.; Balci, S.; Olcum, S.; Süzer, S.; Kocabas, C. Graphene-Based Adaptive Thermal Camouflage. *Nano Lett.* **2018**, *18* (7), 4541–4548. https://doi.org/10.1021/acs.nanolett.8b01746.

(17) Xu, Z.; Li, Q.; Du, K.; Long, S.; Yang, Y.; Cao, X.; Luo, H.; Zhu, H.; Ghosh, P.; Shen, W.; Qiu, M. Spatially Resolved Dynamically Reconfigurable Multilevel Control of Thermal Emission. *Laser Photonics Rev.* **2020**, *14* (1), 1900162. https://doi.org/10.1002/lpor.201900162.

(18) Zhu, H.; Li, Q.; Tao, C.; Hong, Y.; Xu, Z.; Shen, W.; Kaur, S.; Ghosh, P.; Qiu, M. Multispectral Camouflage for Infrared, Visible, Lasers and Microwave with Radiative Cooling. *Nat. Commun.* **2021**, *12* (1), 1805. https://doi.org/10.1038/s41467-021-22051-0.

(19) Hu, R.; Xi, W.; Liu, Y.; Tang, K.; Song, J.; Luo, X.; Wu, J.; Qiu, C.-W. Thermal Camouflaging Metamaterials. *Mater. Today* **2021**, *45*, 120–141. https://doi.org/10.1016/j.mattod.2020.11.013.

(20) Celanovic, I.; Perreault, D.; Kassakian, J. Resonant-Cavity Enhanced Thermal Emission. *Phys. Rev. B* **2005**, *72* (7), 075127. https://doi.org/10.1103/PhysRevB.72.075127.

(21) Ikeda, K.; Miyazaki, H. T.; Kasaya, T.; Yamamoto, K.; Inoue, Y.; Fujimura, K.; Kanakugi, T.; Okada, M.; Hatade, K.; Kitagawa, S. Controlled Thermal Emission of Polarized Infrared Waves from Arrayed Plasmon Nanocavities. *Appl. Phys. Lett.* **2008**, *92* (2), 021117. https://doi.org/10.1063/1.2834903.

(22) Vassant, S.; Moldovan Doyen, I.; Marquier, F.; Pardo, F.; Gennser, U.; Cavanna, A.; Pelouard, J. L.; Greffet, J. J. Electrical Modulation of Emissivity. *Appl. Phys. Lett.* **2013**, *102* (8), 081125. https://doi.org/10.1063/1.4793650.

(23) Brar, V. W.; Jang, M. S.; Sherrott, M.; Lopez, J. J.; Atwater, H. A. Highly Confined Tunable Mid-Infrared Plasmonics in Graphene Nanoresonators. *Nano Lett.* **2013**, *13* (6), 2541–2547. https://doi.org/10.1021/nl400601c.

(24) Brar, V. W.; Sherrott, M. C.; Jang, M. S.; Kim, S.; Kim, L.; Choi, M.; Sweatlock, L. A.; Atwater, H. A. Electronic Modulation of Infrared Radiation in Graphene Plasmonic Resonators. *Nat. Commun.* **2015**, *6* (1), 7032. https://doi.org/10.1038/ncomms8032.

(25) Boriskina, S. V.; Tong, J. K.; Hsu, W.-C.; Liao, B.; Huang, Y.; Chiloyan, V.; Chen, G. Heat Meets Light on the Nanoscale. *Nanophotonics* **2016**, *5* (1), 134–160. https://doi.org/10.1515/nanoph-2016-0010.

(26) Mallawaarachchi, S.; Premaratne, M.; Gunapala, S. D.; Maini, P. K. Tuneable Superradiant Thermal Emitter Assembly. *Phys. Rev. B* **2017**, *95* (15), 155443. https://doi.org/10.1103/PhysRevB.95.155443.

(27) Overvig, A. C.; Mann, S. A.; Alù, A. Thermal Metasurfaces: Complete Emission Control by Combining Local and Nonlocal Light-Matter Interactions. *Phys. Rev. X* **2021**, *11* (2), 021050. https://doi.org/10.1103/PhysRevX.11.021050.

(28) Inoue, T.; Zoysa, M. D.; Asano, T.; Noda, S. Realization of Dynamic Thermal Emission Control. *Nat. Mater.* **2014**, *13* (10), 928–931. https://doi.org/10.1038/nmat4043.

(29) Inoue, T.; De Zoysa, M.; Asano, T.; Noda, S. Electrical Tuning of Emissivity and Linewidth of Thermal Emission Spectra. *Phys. Rev. B* **2015**, *91* (23), 235316. https://doi.org/10.1103/PhysRevB.91.235316.

(30) Coppens, Z. J.; Valentine, J. G. Spatial and Temporal Modulation of Thermal Emission. *Adv. Mater.* **2017**, *29* (39), 1701275. https://doi.org/10.1002/adma.201701275.

(31) Park, J.; Kang, J. H.; Liu, X.; Maddox, S. J.; Tang, K.; McIntyre, P. C.; Bank, S. R.; Brongersma, M. L. Dynamic Thermal Emission Control with InAs-Based Plasmonic Metasurfaces. *Sci. Adv.* **2018**, *4* (12), eaat3163. https://doi.org/10.1126/sciadv.aat3163.

(32) Larciprete, M. C.; Centini, M.; Paoloni, S.; Fratoddi, I.; Dereshgi, S. A.; Tang, K.; Wu, J.; Aydin, K. Adaptive Tuning of Infrared Emission Using VO2 Thin Films. *Sci. Rep.* **2020**, *10* (1), 11544. https://doi.org/10.1038/s41598-020-68334-2.

(33) Liu, X.; Padilla, W. J. Thermochromic Infrared Metamaterials. *Adv. Mater.* **2016**, *28* (5), 871–875. https://doi.org/10.1002/adma.201504525.

(34) Liu, X.; Padilla, W. J. Reconfigurable Room Temperature Metamaterial Infrared Emitter. *Optica* **2017**, *4* (4), 430–433. https://doi.org/10.1364/OPTICA.4.000430.

(35) Horowitz, K. A.; Remo, T. W.; Smith, B.; Ptak, A. J. *A Techno-Economic Analysis and Cost Reduction Roadmap for III-V Solar Cells*; NREL/TP-6A20-72103; National Renewable Energy Lab. (NREL), Golden, CO (United States), 2018. https://doi.org/10.2172/1484349.

(36) Miller, D. A. B.; Zhu, L.; Fan, S.; Pendry, J. B.; Yablonovitch, E. Universal Modal Radiation Laws for All Thermal Emitters. https://doi.org/10.1073/pnas.1701606114.

(37) Lee, B. J.; Zhang, Z. M. Temperature and Doping Dependence of the Radiative Properties of Silicon: Drude Model Revisited. In *2005 13th International Conference on Advanced Thermal Processing of Semiconductors*; 2005; p 10 pp.-. https://doi.org/10.1109/RTP.2005.1613717.

(38) Seitz, F. *The Modern Theory of Solids*; Dover books on physics and chemistry; Dover Publications, 1987.

(39) Baker-Finch, S. C.; McIntosh, K. R.; Yan, D.; Fong, K. C.; Kho, T. C. Near-Infrared Free Carrier Absorption in Heavily Doped Silicon. *J. Appl. Phys.* **2014**, *116* (6), 063106. https://doi.org/10.1063/1.4893176.

(40) Seto, J. Y. W. The Electrical Properties of Polycrystalline Silicon Films. *J. Appl. Phys.* **1975**, *46* (12), 5247–5254. https://doi.org/10.1063/1.321593.

(41) Guo, Y.; Fan, S. Narrowband Thermal Emission from a Uniform Tungsten Surface Critically Coupled with a Photonic Crystal Guided Resonance. *Opt. Express* **2016**, *24* (26), 29896–29907. https://doi.org/10.1364/OE.24.029896.

(42) Fan, S.; Joannopoulos, J. D. Analysis of Guided Resonances in Photonic Crystal Slabs. *Phys. Rev. B* **2002**, *65* (23), 235112. https://doi.org/10.1103/PhysRevB.65.235112.

(43) Liu, V.; Fan, S. S 4 : A Free Electromagnetic Solver for Layered Periodic Structures. *Comput. Phys. Commun.* **2012**, *183*, 2233–2244. https://doi.org/10.1016/j.cpc.2012.04.026.

(44) Song, A. Y.; Catrysse, P. B.; Fan, S. Broadband Control of Topological Nodes in Electromagnetic Fields. *Phys. Rev. Lett.* **2018**, *120* (19), 193903. https://doi.org/10.1103/PhysRevLett.120.193903.

(45) Sze, S. M.; Li, Y.; Ng, K. K. *Physics of Semiconductor Devices*; Wiley, 2021.

(46) Sun, K.; Cai, Y.; Huang, L.; Han, Z. Ultra-Narrowband and Rainbow-Free Mid-Infrared Thermal Emitters Enabled by a Flat Band Design in Distorted Photonic Lattices. *Nat. Commun.* **2024**, *15* (1), 4019. https://doi.org/10.1038/s41467-024-48499-4.

(47) Liu, X.; Qiao, Q.; Dong, B.; Liu, W.; Xu, C.; Xu, S.; Zhou, G. MEMS Enabled Suspended Silicon Waveguide Platform for Long-Wave Infrared Modulation Applications. *Int. J. Optomechatronics* **2022**, *16* (1), 42–57. https://doi.org/10.1080/15599612.2022.2137608.

(48) Su, Y.; Zhang, Y.; Qiu, C.; Guo, X.; Sun, L. Silicon Photonic Platform for Passive Waveguide Devices: Materials, Fabrication, and Applications. *Adv. Mater. Technol.* **2020**, *5* (8), 1901153. https://doi.org/10.1002/admt.201901153.